\documentclass[fleqn,usenatbib]{mnras}

\usepackage{newtxtext,newtxmath}

\usepackage[T1]{fontenc}

\DeclareRobustCommand{\VAN}[3]{#2}
\let\VANthebibliography\thebibliography
\def\thebibliography{\DeclareRobustCommand{\VAN}[3]{##3}\VANthebibliography}

\usepackage{graphicx}	
\usepackage{amsmath}	
\usepackage{lineno}

\defcitealias{Wiseman2022}{W22}
\defcitealias{Briday2021}{B22}

\title[Galaxy-age-driven SN Ia luminosity steps]{Further evidence that galaxy age drives observed type Ia supernova luminosity differences}

\author[P. Wiseman et al.]{
P. Wiseman,$^{1}$\thanks{E-mail: p.s.wiseman@soton.ac.uk (PW)}
M. Sullivan,$^{1}$
M. Smith$^{2}$
and B. Popovic$^{3}$
\\
$^{1}$ School of Physics and Astronomy, University of Southampton, Southampton, SO17 1BJ, UK\\
$^{2}$ Univ Lyon, Univ Claude Bernard Lyon 1, CNRS, IP2I Lyon / IN2P3, IMR 5822, F-69622 Villeurbanne, France\\
$^{3}$ Department of Physics, Duke University, Durham, NC 27708, USA\\
}

\date{Accepted XXX. Received YYY; in original form ZZZ}

\pubyear{2023}

\begin{document}
\label{firstpage}
\pagerange{\pageref{firstpage}--\pageref{lastpage}}
\maketitle

\begin{abstract}
Type Ia supernovae (SNe Ia) are explosions of white dwarf stars that facilitate exquisite measurements of cosmological expansion history, but improvements in accuracy and precision are hindered by observational biases. Of particular concern is the apparent difference in the corrected brightnesses of SNe Ia in different host galaxy environments. SNe Ia in more massive, passive, older environments appear brighter after having been standardized by their light-curve properties. The luminosity difference commonly takes the form of a step function. Recent works imply that environmental characteristics that trace the age of the stellar population in the vicinity of SNe show the largest steps. Here we use simulations of SN Ia populations to test the impact of using different tracers and investigate promising new models of the step. We test models with a total-to-selective dust extinction ratio $R_V$ that changes between young and old SN Ia host galaxies, as well as an intrinsic luminosity difference between SNe from young and old progenitors. The data are well replicated by a model driven by a galaxy-age varying $R_V$ and no intrinsic SN luminosity difference, and we find that specific star-formation rate measured locally to the SN is a relatively pure tracer of this galaxy age difference. We cannot rule out an intrinsic difference causing part of the observed step and show that if luminosity differences are caused by multiple drivers then no single environmental measurement is able to accurately trace them. We encourage the use of multiple tracers in luminosity corrections to negate this issue.
\end{abstract}



\section{Introduction}

Type Ia supernovae (SNe Ia) are routinely used as standardizable candles to directly probe the expansion history of the Universe, a technique that led to the discovery of the accelerated cosmic expansion \citep{Riess1998,Perlmutter1999}. These measurements are facilitated by empirical relationships between the optical brightness of a SN Ia and its decline rate (commonly known as `stretch’; \citealt{Phillips1993}) and a SN Ia brightness and its colour \citep{Riess1996,Tripp1998}. When combined, these two relationships can reduce the dispersion in SN Ia distances to 7 per cent ($\simeq0.15$~mag), enabling modern analyses to measure the dark energy equation-of-state parameter to a precision of two per cent \citep{Scolnic2018,Abbott2019,Brout2022}. Key to achieving an accuracy that matches this precision is the comprehensive modelling of both SN Ia selection biases in large surveys \citep[e.g.][]{Kessler2009,Perrett2010,Betoule2014,Kessler2019,Vincenzi2020} and astrophysical processes affecting the observed SN Ia properties. 

In the latter case, significant effort has been dedicated to explaining and accounting for the observed relationships between the standardized (i.e., stretch- and colour-corrected) SN Ia luminosities and their host galaxy parameters, such as stellar mass \citep{Kelly2010,Lampeitl2010,Sullivan2010}, star-formation rate (SFR; \citealt{Sullivan2010}), morphology \citep{Sullivan2003,Hakobyan2020}, stellar age \citep{Rose2019,Rose2021,Millan-Irigoyen2022}, metallicity \citep{Gallagher2008,Childress2013a,Moreno-Raya2016,Millan-Irigoyen2022}, or galaxy rest-frame colour \citep{Roman2018, Jones2018, Kelsey2021,Kelsey2022}. The relationships usually take the form of a \lq step\rq, where the mean standardized luminosities of SNe Ia on different sides of a threshold in a given host galaxy parameter (e.g., a stellar mass of $\log(M_*/\mathrm{M}_{\odot})=10$) differ by a fixed amount.

SN Ia stretch also correlates with various global (i.e., measurements of the entire galaxy) host galaxy parameters \citep[e.g.,][]{Hamuy1995,Sullivan2006}, and in particular those that trace stellar population age. \citet{Rigault2013}, \citet{Rigault2018} and \citet{Nicolas2020} show evidence that there are at least two modes of SN stretch,one of which is dominant for young white dwarfs and the other in older progenitor systems. There is also evidence that the luminosity step is more significant when computed for a `locally’-measured\footnote{Measured in a small aperture centered at the SN location.} property \citep{Rigault2018,Roman2018,Rose2019,Kelsey2021,Kelsey2022}, particularly properties linked to age, suggesting that SNe Ia belonging to each stretch population may also have differing mean intrinsic luminosities. \citet[][hereafter B22]{Briday2021} showed that the magnitude of the step is directly related to the ability of a given host galaxy measurement to trace the age of the stellar population near the SN: locally measured specific SFR (the SFR normalised by stellar mass; \citealt{Guzman1997}) displays the largest step, while the step in global galaxy morphology (which has a far less direct connection to the stellar population age at the SN Ia position) is smallest. \citetalias{Briday2021} propose that there should be a linear relationship between the step size and how directly the host galaxy parameter traces local sSFR, and their observed data agree with the prediction.

An alternative explanation is that the stellar mass step is a result of different average dust properties in low- and high-mass galaxies \citep{Brout2020}. For example, a difference in the mean total-to-selective extinction ratio ($R_V$) such that mean $R_V =2.0$ in low-mass galaxies and $R_V =3.0$ in high-mass galaxies can reduce the mass step to $\sim0.02$\,mag \citep{Popovic2021a}.
Such a difference in $R_V$ can explain the observed difference in the colour--luminosity coefficient $\beta$ between SNe Ia in low and high mass host galaxies \citep{Sullivan2011,Gonzalez-Gaitan2020,Chen2022,Kelsey2022}, and also account for the observed increase in the step size as a function of SN colour \citep{Brout2020,Popovic2021a,Kelsey2022} as well as the increase in SN Ia luminosity dispersion as a function of SN colour \citep{Brout2020,Popovic2021a}.

\citet[][hereafter W22]{Wiseman2022} explored variations of the \citet{Brout2020} model by simulating the stellar age distributions of galaxies and the delay-time distributions (DTDs) of SNe Ia. The DTD is the probability distribution describing the \lq delay\rq\ between the star formation event in which the SN progenitor was formed and the epoch when the SN progenitor explodes. This is commonly modelled as a power law of form $t^{\lambda}$ \citep[e.g.,][]{Maoz2014}\footnote{Usually the power-law index is represented by $\beta$. Here we use $\lambda$ to avoid confusion with the SN colour -- luminosity relation.}, with $\lambda\sim-1$ favoured by observations. By incorporating the DTD with the evolution of stellar ages within simulated galaxies, \citetalias{Wiseman2022} trace the connection between the intrinsic properties of stellar age and SN Ia progenitor age, with measurable galaxy properties such as stellar mass, SFR, and rest-frame colour, and how these link to SN Ia distance measurements.

\citetalias{Wiseman2022} find that model with a mean $R_V$ of a SN Ia population ($\overline{R_V}$) that changes with galaxy age (rather than stellar mass) best reproduces the observed step and its evolution with SN colours -- in particular when this model is compared to observations of both host galaxy stellar mass and host galaxy colour. This is a complex, multi-dimensional parameter space and several permutations of the model can explain the broad trends in the data equally well, although no model is able to reproduce the data in both host parameters simultaneously. In particular, models with and without an intrinsic SN Ia luminosity step perform equally well, with different observed luminosity step sizes accounted for by varying differences in the $R_V$ population means. A model where $R_V$ changes with respect to the age of the SN Ia progenitor itself (e.g., the extinction is caused by circumstellar material that is affected by the astrophysics of the white dwarf and its companion) does not provide good fits to the data.

Other analyses have found conflicting results. By measuring $R_V$ for individual SNe Ia using a hierarchical Bayesian model, \citet{Thorp2021} and \citet{Thorp2022} found no significant difference in the $R_V$ population means between SNe in low and high mass galaxies. There is also some evidence that the step remains when measuring distances using near infra-red (NIR) light curves, where the effects of dust should have a smaller impact \citep{Ponder2020,Uddin2020, Jones2022}, although, by contrast, \citet{Johansson2021} found the step was removed in both the NIR and when fitting each SN Ia for its $R_V$.

Evidence for a large difference between the dust laws in the general population of old/massive galaxies and young/low-mass galaxies is also limited, hindered by the difficulty in measuring dust laws for large samples of galaxies typically only observed with optical and sometimes NIR photometry. \citet{Salim2018} showed that the $R_V$ values for \textit{star-forming} galaxies \textit{increase} with stellar mass, opposite to the inference of the \citet{Brout2020} model. This increase corresponds to an increase in the dispersion of the $R_V$ parameter and an increase in the optical depth (which correlates with stellar mass). However, \citet{Salim2018} also show that the $R_V$ of \textit{passive} galaxies is $\sim 0.5-1$ smaller than star-forming galaxies, indicating that the dust is dependent on age/star-formation history in the direction matching the \citetalias{Wiseman2022} model. A similar effect has been observed in SN Ia host galaxies by \citet{Meldorf2022} and \citet{Duarte2022}. The increasing fraction of passive galaxies (and also of passive SN Ia hosts) with stellar mass may thus explain the inferred decrease in $R_V$ with stellar mass inferred by \citet{Brout2020}, \citet{Popovic2021} and \citetalias{Wiseman2022}.

Overall, dust-based models can explain the optical SN Ia data, but it is unclear which host galaxy parameter is the key driver of a changing dust law and whether dust can account for the full step. The tests conducted by \citetalias{Briday2021} provide a framework to disentangle the problem: a successful model must reproduce both the trend of luminosity step as a function of SN Ia colour as well as the relationship between step magnitude and tracer contamination. 

In this work, we test the dust-based models explored in \citetalias{Wiseman2022} in the framework of \citetalias{Briday2021}, which we outline in Section \ref{sec:method}. Using a simulated sample of SNe Ia, we assume each potential astrophysical driver of a changing dust law as a `truth’ and assess the resulting relationship between step size and contamination between the two SN Ia populations traced by each host galaxy property, which we present in Section \ref{sec:results}. In Sections \ref{sec:discussion}  and \ref{sec:conclusions} we asses and review the implications of our findings. Where appropriate, we assume a reference cosmology described by a spatially flat $\Lambda$CDM model with $H_0 = 70~ \mathrm{km}~\mathrm{s}^{-1}~\mathrm{Mpc}^{-1}$ and $\Omega_M = 0.3$.

\begin{table*}
 \caption{SN Ia luminosity models tested. $Y$ is the environmental property on which $R_V$ varies between $\overline{R_{V,1}}$ and $\overline{R_{V,2}}$. $X$ is the astrophysical driver of additional luminosity steps that have magnitude $\gamma_X$.}
 \label{tab:models}
 \begin{tabular}{lccccccccc} 
		\hline
		Model Name & $Y$ & $\overline{R_{V,1}}$   &  $\overline{R_{V,2}}$    &  $X$  & $\gamma_{X}$  \\
		\hline
		SN step only & -  & 2.5 & 2.5 & SN progenitor age &0.2\\
	    Age $R_V$& Host stellar age  & 1.5 & 3.0 & - &0.0\\
        Age $R_V~+$ SN step &  Host stellar age & 1.75 & 2.5  &SN progenitor age&  0.15\\
       	\hline
		
	\end{tabular}
\end{table*}

\begin{figure*}

	\includegraphics[width=\textwidth]{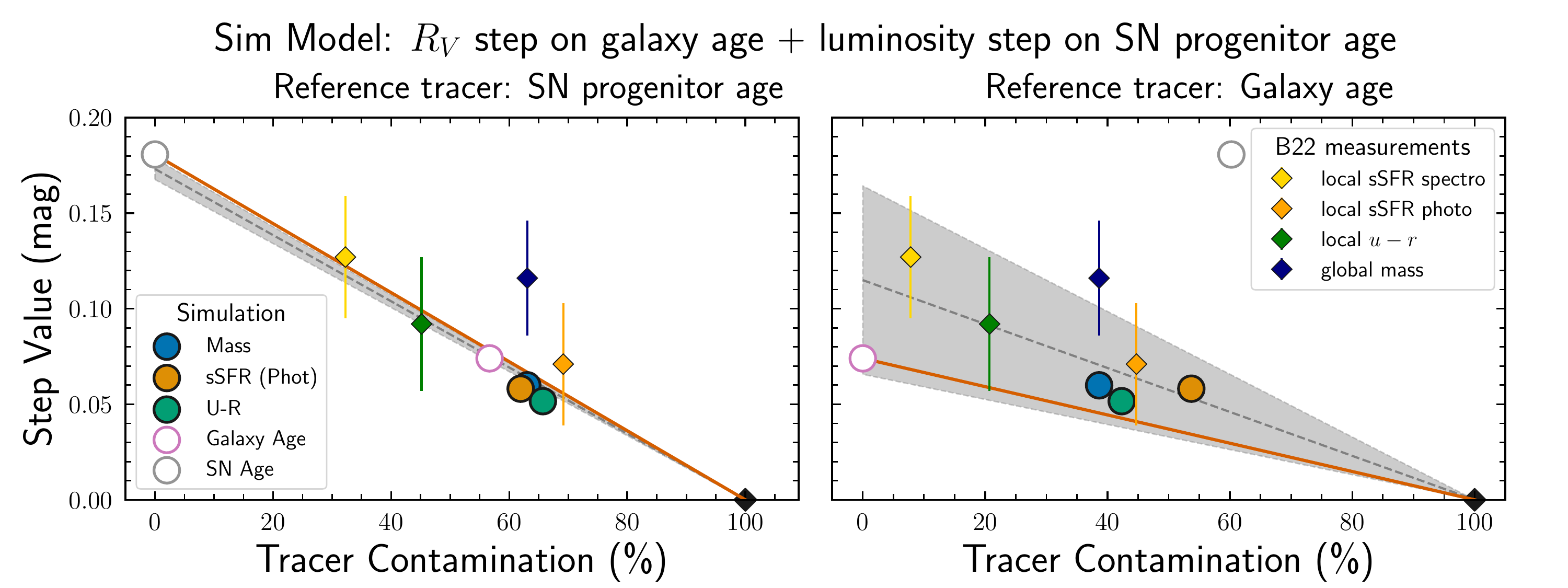}
    \caption{The measured SN Ia luminosity step $\gamma$ (Section~\ref{subsec:sndistances}) as a function of the contamination of environmental tracers, using two different reference tracers. The simulations have a constant $R_V$ and a $0.2$\,mag luminosity step on SN progenitor age. The solid line (the \lq model\rq) illustrates a direct relationship between the true step at zero contamination and a zero step at 100 per cent contamination. Circles correspond to the simulation from this work, diamonds are observational measurements taken from \citetalias{Briday2021}. Open circles are intrinsic galaxy properties taken from the simulation, while filled markers are galaxy properties estimated from spectral energy distribution fitting (either of data or of simulations outputs). Simulated properties are shown without uncertainties since in the simulation they are known exactly. The dashed line and shaded region show a least-squares fit to the simulated steps (forced through the point [100\%, 0]) and $1$\,$\sigma$ uncertainties respectively. \textit{Left}: contamination computed assuming SN progenitor age as the reference tracer; \textit{Right}: contamination computed incorrectly assuming galaxy age as the reference tracer.}
    \label{fig:SNage_step_only}
\end{figure*}

\section{Method}
\label{sec:method}

In \citetalias{Wiseman2022} a framework was introduced for tracing the effects of galaxy-scale astrophysical effects through to SN Ia distance measurements in a simulation. The simulation comprises two core components described in detail in previous works and summarised in this section: firstly, we generate a library of potential host galaxies \citep{Wiseman2021}, and then we generate samples of SN Ia light-curve parameters associated with those galaxies (\citetalias{Wiseman2022}). The simulations are optimised to best reproduce SN samples from the Dark Energy Survey supernova program (DES-SN; \citealt{Bernstein2012}), but in principle they can be adapted to mimic any SN survey. Unless otherwise stated, we use identical model parameters to \citetalias{Wiseman2022}, shown in Table \ref{tab:models}.

\subsection{Galaxy models}

To track stellar populations through the evolution of galaxies, we use the galaxy mass assembly prescriptions of \citet{Wiseman2021} based upon the work of \citet{Childress2014}. These simulations begin with seed galaxies at a range of look-back times and follow them as they evolve along the star-forming main sequence according to an empirical relation between stellar mass, SFR, and redshift. The simulations also include the quenching of star formation. Here we follow the updated prescription of \citetalias{Wiseman2022}, where each seed galaxy is replicated 100 times, with the time of quenching onset drawn from a probability distribution. Small bursts of star formation happening at random times are drawn in a similar fashion. The output of the simulation is a representative library of galaxies across a large range of stellar mass and redshift, for which the full stellar age distribution is known at a resolution of 0.5\,Myr. This star-formation history for each galaxy is convolved with a SN Ia DTD to produce a probability distribution for the age of SN Ia progenitors at the time of explosion. We use the power-law DTD from \citet{Wiseman2021} with $\lambda=-1.13$. 

\subsection{Supernova models}
\label{subsec:models}

SNe Ia in the simulation are represented by their light-curve parameters and corresponding uncertainties. We parametrise light curves in the SALT framework \citep{Guy2007}, with SN colour represented by the parameter $c$ and SN stretch by the parameter $x_1$. We assign intrinsic $c$ and $x_1$ to each SN by drawing from probability distributions that relate to their host galaxy or their progenitor age. In this work we use the two-component colour distribution of \citet{Brout2020}, with an intrinsic component following a Gaussian distribution and an external reddening component drawn from an exponential distribution. We draw $x_1$ from a Gaussian mixture introduced by \citet{Nicolas2020}, with the component weights determined by SN progenitor age, using identical parameters to \citetalias{Wiseman2022}.

In cosmological SN Ia samples, SN absolute magnitudes show scatter around their mean. Various models have been introduced in order to account for this scatter, which are summarised in \citet{Brout2020} and \citet{Popovic2021}. In the \citet{Brout2020} framework used in this work, the intrinsic scatter is caused by SN to SN variation in the intrinsic colour-luminosity coefficient $\beta_{\mathrm{SN}}$ and dust extinction $R_V$. In our simulations, the SN Ia intrinsic luminosity $M_B$ is fixed to $-19.365$\,mag. This intrinsic brightness is modified according to the intrinsic stretch using a linear relationship with a coefficient fixed to $\alpha_{\mathrm{SN}}=0.15$, and by the intrinsic colour--luminosity relationship drawing the coefficient $\beta_{\mathrm{SN}}$ from a Gaussian distribution with mean 2.0 and standard deviation 0.35, which were the best-fitting values for the scatter model in \citet{Brout2020}. This adjusted absolute magnitude is then converted to an apparent magnitude $m_B$ using a distance modulus computed from a fixed reference cosmology. The brightness is reduced by $\Delta m_B$, which is determined according to the extinction via
\begin{equation}
    \Delta m_B = (R_V +1)\times E(B-V)\,,
    \label{eq:deltamb}
\end{equation}
where $R_V$ is the total-to-selective extinction ratio and reddening $E(B-V)$ is drawn from an exponential distribution. 

The \citetalias{Wiseman2022} models draw the $R_V$ value for a given SN Ia from one of two Gaussian distributions with different means but identical standard deviations, with the means depending on a given host galaxy parameter $Y$: those SNe Ia hosted in galaxies above a specific value of $Y$ are drawn from a distribution with a different mean than those SNe Ia hosted in galaxies below that value of $Y$. In this work, we focus only on using the host galaxy stellar-mass-weighted mean stellar age (which we refer to simply as stellar age) for $Y$ as that was the best matching parameter in \citetalias{Wiseman2022}.
Additionally, we allow for an intrinsic luminosity step driven by different host properties/progenitor ages that is \textit{not} related to $R_V$ differences: this luminosity step $\gamma_{X,\mathrm{sim}}$ is added or subtracted to the distance modulus of the SNe Ia directly, rather than as a modification to the peak brightness $m_B$. Here, we use the SN progenitor age for $X$. A summary of the models we test is presented in Table~\ref{tab:models}.

We simulate 2000 SNe Ia in the redshift range $0\leq z \leq 1.2$. To mimic the effect of observational survey selection effects, we apply efficiency corrections following the DES host galaxy redshift detection efficiency as modelled by \citet{Vincenzi2020}, such that SNe in fainter host galaxies are less likely to be included in the final sample. We add Gaussian noise to each of $m_B$, $c$ and $x_1$ according to the SN Ia brightness and redshift using empirical relations measured from DES-SN data (\citetalias{Wiseman2022}).

\subsection{Supernova distances and residual steps}
\label{subsec:sndistances}

Using the `observed' parameters modelled in the previous section, distance moduli $\mu_{\mathrm{obs}}$ of the simulated SNe are computed using a form of the Tripp estimator:
\begin{equation}
    \mu_{\mathrm{obs}}= m_B - M_B + \alpha x_1 - \beta c + \gamma_{X,\mathrm{obs}} \Gamma_X \,,
\label{eq:tripp}
\end{equation}
where $\alpha$, $\beta$\footnote{Note that $\beta$ here accounts for all colour--luminosity dependence, different from $\beta_{\mathrm{SN}}$ in the simulations which only corresponds to intrinsic variation and not dust.}, $\gamma_{X,\mathrm{obs}}$, $M_B$ are nuisance parameters, and $\Gamma_X$ represents the step for some environmental property $X$:
\begin{equation}
 \Gamma_X = \left\{
    \begin{array}{@{}rc@{}}
    0.5, & X < X_{\mathrm{split}} \\
    -0.5 , & X \geq X_{\mathrm{split}} \\
    \end{array}\right. \,,
        \label{eq:Gamma}
\end{equation} 
where $X_{\mathrm{split}}$ is the location of the step for that property. To find the values of the nuisance parameters $\alpha$, $\beta$ and $\gamma_X$, we minimize the $\chi^2$ statistic for the observed distance moduli $\mu_{\mathrm{obs}}$ compared to the fixed cosmological model $\mu_{\mathrm{mod}}$, given the distance uncertainties and their covariance.  

\subsection{Step drivers, tracers, and contamination}
To measure the accuracy of environmental tracers of SN Ia luminosity variations we follow the method of \citetalias{Briday2021}. This prescription is based on the assumption that SNe Ia are divided into two populations ($a$ and $b$) that have a dependency on some physical property that we denote the \textit{driver}. Different environmental measurements, \textit{tracers}, map to those populations with differing levels of accuracy: a tracer that is closely related to the true driver of the population split will accurately divide the populations, while a tracer less closely linked to the SN populations will not split the populations well. The degree of accuracy with which an environmental tracer recovers the true driver of the population split is parametrised by \textit{contamination} ($C$), which is defined as the fraction of SNe Ia that are classified as belonging to population $a$, but truly belong to $b$, plus the opposite: SNe classified as $b$ but truly $a$, which we define briefly below. 

For a given tracer, $N_a$ represents objects that truly belong to the class $a$ while $N^a$ is the number of objects classified as $a$. Thus the contamination can be defined as the fraction of objects classified as belonging to $a$ that truly belong to $b$: $C^a\equiv N^a_b/N^a$. Similarly there is contamination $C_b$: objects classified as $b$ but truly $a$: $C^b\equiv N^b_a/N^b$. Total contamination $C$ is then defined as $C\equiv C^a + C^b$.  A tracer that perfectly separates the two SN populations has $C=0$, while a random tracer carrying no information about the SN populations has $C=100$ per cent. 

The framework presented in \citetalias{Briday2021} predicts that there is a direct relationship between the measured step size and contamination: a tracer with no contamination has the largest step, while that with 100 per cent contamination has no step. Observations should lie directly on the line that connects these points but are dispersed by observational noise. We call this line the \citetalias{Briday2021} \textit{step -- contamination relation}. While the true driver is not known astrophysically, \citetalias{Briday2021} define a \textit{reference tracer} to split the SNe, for which the contamination is defined to be zero. Hypothesising that the SN Ia populations are separated by their progenitor age, they use spectroscopic local sSFR as a reference tracer, against which all other tracers display some contamination. Local sSFR itself may be -- and probably is -- an imperfect tracer: therefore against the perfect reference tracer, local sSFR will also show some contamination.

In this analysis, there are two key differences to the work of \citetalias{Briday2021}. Firstly, for each simulation, we know the driver(s) of the SN Ia luminosity steps in that simulation. For simulations where these steps are caused by a single driver, we can use that driver as a reference tracer knowing it has zero contamination, but also test other reference tracers to simulate the real-world effects of choosing an incorrect reference tracer.

Secondly, not only can we use observable properties such as stellar mass, galaxy colours, and sSFR as tracers, but we can also use intrinsic properties such as SN Ia progenitor age and galaxy age, which cannot be observed directly. Thus, by simulating SN Ia samples with luminosity steps driven by SN Ia progenitor age and galaxy age, measuring their steps and contamination from observable properties, and comparing to the model of \citetalias{Briday2021}, we can break the degeneracies between the models of \citetalias{Wiseman2022}. In our analysis we investigate the following tracers: global host-galaxy rest-frame $U-R$ colour, global host-galaxy sSFR (as would be measured by a photometric spectral energy distribution fitting method), and the global host-galaxy stellar mass. Note that our simulations are not resolved, i.e., our simulations cannot estimate local tracers (e.g., local sSFR).

To compare step sizes measured in data to those in our simulation, we include the data used by \citetalias{Briday2021}. Our simulations do not provide the measurements of local sSFR that \citetalias{Briday2021} use as a reference tracer in the data, so we calibrate the contamination measurements by fixing the measured contamination of global stellar mass from \citetalias{Briday2021} to be the same as in our simulation, and use the relative differences between the contamination of stellar mass and the other tracers used by \citetalias{Briday2021} to include in our analysis. The \citetalias{Briday2021} contaminations and steps were measured on data, and so the relative differences are only valid given the true underlying model. However, we show them against all of our simulations: the agreement between our simulated and the measured contamination and steps for each tracer and in each model provides additional information on which model is most realistic.

\begin{figure*}

	\includegraphics[width=\textwidth]{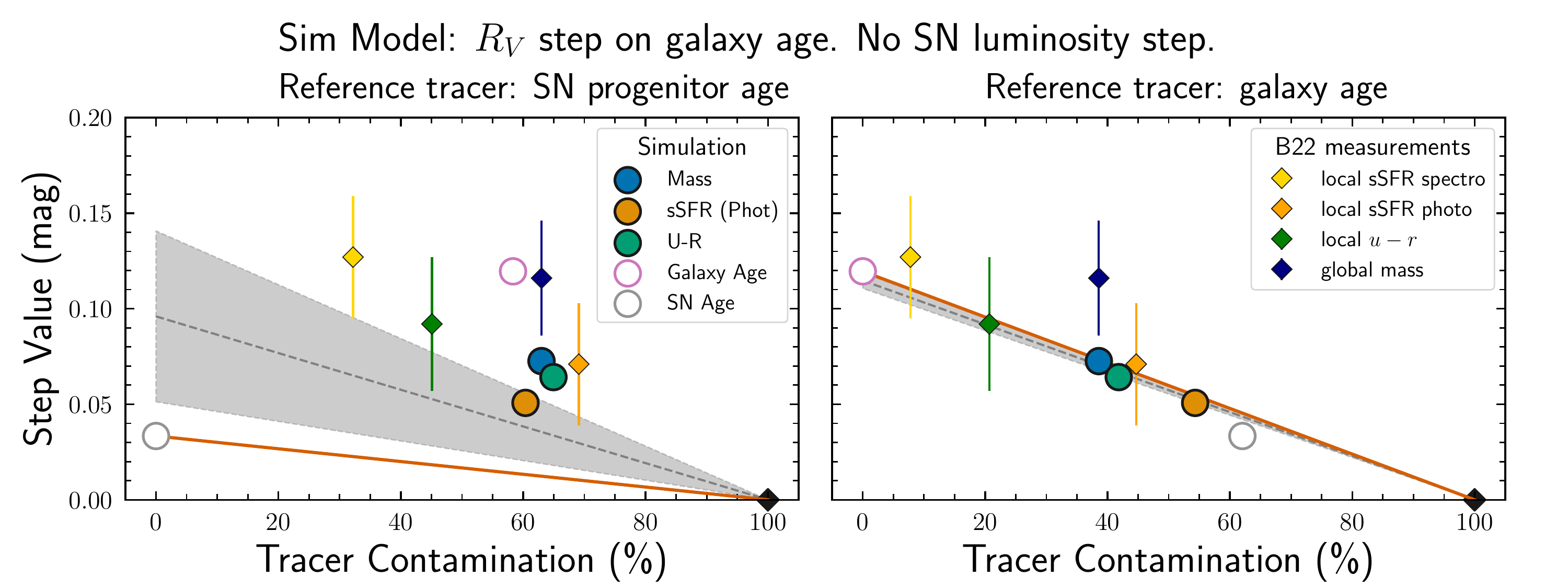}
    \caption{The SN Ia luminosity step as a function of the contamination of environmental tracers for the model where galaxy age is the astrophysical driver: an $R_V$ difference of 1.5 between old and young galaxies and no intrinsic luminosity step on SN Ia progenitor age. 
    \textit{Left}: contamination computed assuming, incorrectly, SN Ia progenitor age as the reference tracer; \textit{Right}: contamination computed correctly assuming galaxy age as the reference tracer.}
    \label{fig:Rv_only}
\end{figure*}

\begin{figure*}

	\includegraphics[width=\textwidth]{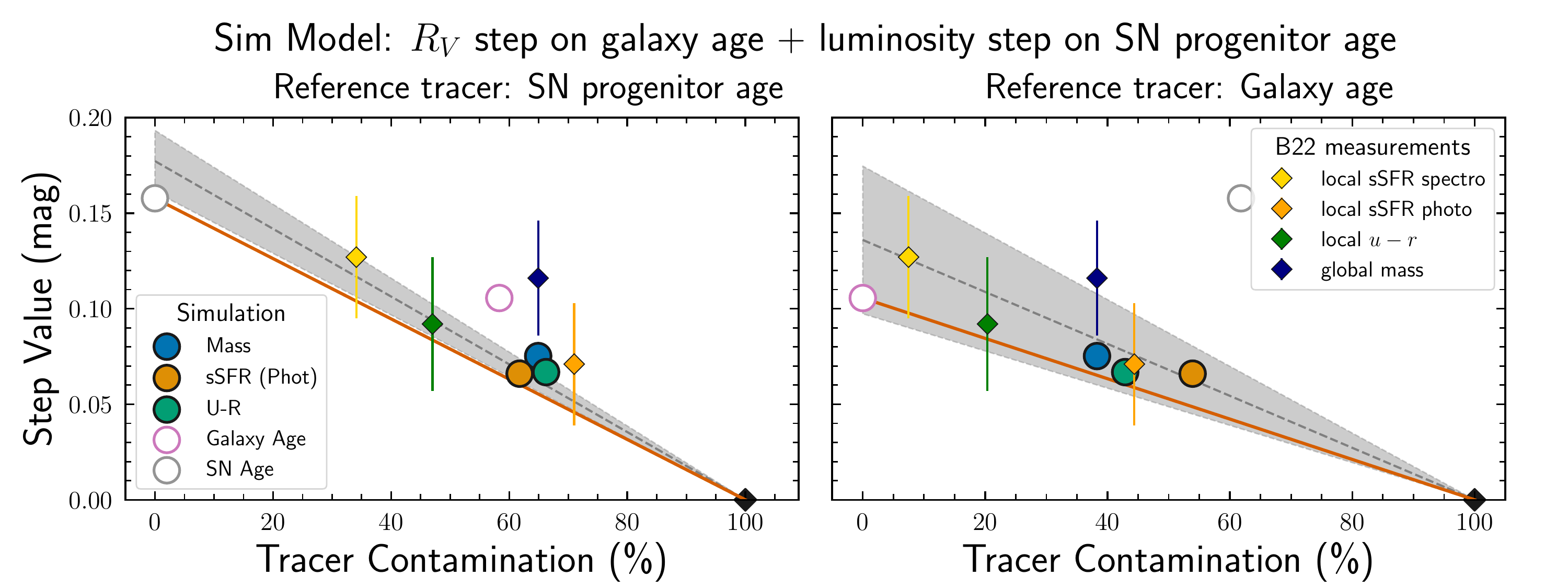}
    \caption{The same as Fig.~\ref{fig:Rv_only}, but where galaxy age and SN progenitor age are both astrophysical drivers: an $R_V$ difference of 0.75 between old and young galaxies and a 0.15~mag intrinsic luminosity step on SN Ia progenitor age. \textit{Left}: contamination computed assuming SN Ia progenitor age as the reference tracer; \textit{Right}: contamination computed assuming galaxy age as the reference tracer.}
    \label{fig:Rv+SNage}
\end{figure*}

\section{Results}
\label{sec:results}
 
\citetalias{Briday2021} demonstrated that, under the assumption that SNe Ia can be split into two populations with different mean luminosities, there is a linear relationship between the ability of an environmental tracer to divide the SNe between those two populations and the size of the recovered luminosity step. In that analysis, low-redshift samples of SNe Ia followed the step--contamination relationship when using local sSFR as a reference tracer. Any model of SN Ia populations that seeks to explain the luminosity step should display a similar trend to that observed in the data. We validate this method by testing a simulation with a simple luminosity step on SN Ia age (Section~\ref{subsec:res_age_step_only}). We then construct step--contamination relationships for the two best-matching models to the step versus SN Ia colour relationships from \citetalias{Wiseman2022} in Section \ref{subsec:res_age_rv_steps}.
 
\subsection{SN Ia progenitor age step only}
\label{subsec:res_age_step_only}

We begin with a simulation with fixed $R_V$ and a $0.2$\,mag step in the intrinsic SN Ia luminosity at a progenitor age of $0.75$\,Gyr. We then use the progenitor age as a reference tracer -- in the simulation this is the true driver, so it has a true contamination of zero. The contamination and step measurements measured for the other tracers are shown in the left-hand panel of Fig.~\ref{fig:SNage_step_only}. The $0.2$\,mag step is almost fully recovered when fitting for $\gamma_{\mathrm{SNage}}$ -- a slight reduction in the step is caused by SN Ia progenitor age affecting both $\gamma$ and $x_1$ simultaneously.

The other environmental tracers are found at higher contaminations and lower step sizes, as expected from the \citetalias{Briday2021} step -- contamination relationship (shown as a solid line). We show a least-squares regression line between the points (the dashed line and shaded uncertainty), forcing the line through the point of [100\% contamination, zero step]. The small uncertainty in this best-fitting line illustrates the validity of using progenitor age as the reference tracer given this SN model, and the \citetalias{Briday2021} measurements lying close to the predicted step -- contamination line provides reassurance that our ad-hoc connection between the simulated and observed measurements is reasonable. The exception is stellar mass, which shows a large discrepancy between the observed and simulated step size. For the correct model, the recovered contamination and step size should be consistent between observations and the simulation. The implication is thus that an SN-progenitor-age -- luminosity step does not explain the full observed effect. Indications that this model is not a full description of the effect are also apparent in step size versus SN colour relationships analysed \citet{Kelsey2022} and \citetalias{Wiseman2022}.

Our second test is to incorrectly use galaxy age as a reference tracer for the simulation that is driven by an SN Ia progenitor-age step (Fig.~\ref{fig:SNage_step_only}, right). As expected, the agreement is worse than when using SN Ia progenitor age as the reference tracer. SN progenitor age has the largest contamination (60 per cent) but (as expected) the largest step. The points all lie above the \citetalias{Briday2021} step -- contamination relationship, indicating that the chosen reference tracer is not the true driver: the actual contamination of that tracer should be non-zero, which would increase the gradient of the step -- contamination line and allow it to be more consistent with the other points. The observed local measurements of spectroscopic sSFR and $u-r$ colour show less contamination here than when SN progenitor age is used as a reference tracer. These lower contamination measurements, along with the high contamination for SN progenitor age, imply that local spectroscopic sSFR and local $u-r$ are more closely related to galaxy age than they are to SN progenitor age. This is a consequence of the SN Ia DTD which decouples the observed stellar population age with that of SN progenitors: roughly half of SNe Ia explode over 1\,Gyr since the star formation event that formed the progenitor star.

There are some notable discrepancies between the observed and simulated contaminations, whichever reference tracer is used. A clear example is that the contamination for galaxy colour is much larger in the simulations than in the observations. Part of this difference can be attributed to the simulations only tracing global galaxy colour while the observations are measured locally to the SN: the inability of our simulations to predict locally-measured properties is a limitation that will be rectified with future developments.

A second discrepancy is the relative lack of dynamic range in the contamination of simulated tracers, whereas the contaminations of observed tracers show greater dispersion. For example, using SN progenitor age as the reference tracer, the other simulated tracers show a range of $55$--$70$ per cent contamination; using galaxy age as the reference tracer the other simulated tracers cover a $40$--$60$ per cent range. A specific example is that both stellar mass and $U-R$ colour (both routinely used in the literature) are $\sim60$ per cent contaminated with SN progenitor age as the reference tracer, and around $40$ per cent contaminated with galaxy age as the reference tracer. This results from the strong covariance between SNe in calculating the contamination of these tracers. For example, when using galaxy age as a reference tracer, 89 per cent of the SNe misclassified by stellar mass are also misclassified by $U-R$ colour. This strong correlation between the simulated tracers may indicate oversimplifications in the galaxy evolution model, such as the nature of starbursts, the quenching of star formation, ultraviolet ionising continuum, dust attenuation, or active galactic nucleus activity.

\subsection{Galaxy age, dust, and SN progenitor age steps}
\label{subsec:res_age_rv_steps}

Having introduced the method using a single driver (a fixed luminosity step driven by SN Ia progenitor age), we next investigate models that have previously been shown to best reproduce the variation of SN distance moduli as a function of host and SN properties. \citetalias{Wiseman2022} showed that a SN progenitor-age --luminosity step alone cannot explain the observed trend of step size increasing with SN colour. We therefore introduce the two successful models from that paper that are built upon a dust law that varies with galaxy age, as introduced in Section \ref{subsec:models}. First, we use only that varying dust law (Section \ref{subsubsec:res_age_rv_step_only}), and then we combine it with the intrinsic SN progenitor-age -- luminosity step (Section \ref{subsubsec:res_ageplusstep}).

\subsubsection{Galaxy age $R_V$ step only}
\label{subsubsec:res_age_rv_step_only}

Fig. \ref{fig:Rv_only} shows the relationships between step size and contamination for a model with a large $\overline{R_V}$ difference of 1.5 between old and young host galaxies, and no intrinsic SN luminosity step. As expected, using SN progenitor age as a reference tracer (left panel) does not result in a good agreement between prediction and either simulations or data: all of the points lie above the \citetalias{Briday2021} step -- contamination relationship. Assuming galaxy age as the reference tracer performs much better (right panel): the dispersion on the best-fitting line is small and in agreement with the model prediction. 

As also expected, local spectroscopic sSFR displays the largest step, as found by \citet{Rigault2018} and \citetalias{Briday2021}. Interestingly, SN progenitor age is the most contaminated and has the smallest step of all tracers tested. As with the model with the SN progenitor-age step only, the simulated host stellar mass step is smaller than the observed one, meaning that the model does not account for the full luminosity difference observed in galaxies of different masses, despite adequately reproducing the step sizes for the other tracers considered.

\subsubsection{Galaxy age $R_V$ step plus SN progenitor-age -- luminosity step}
\label{subsubsec:res_ageplusstep}

In Fig. \ref{fig:Rv+SNage} we show a model where the difference between $R_V$ in young and old galaxies is smaller, and an additional intrinsic luminosity step as a function of SN progenitor age is added. With this model, neither using SN progenitor age nor galaxy age as reference tracers results in good agreement between the step -- contamination prediction (solid line) and the data/simulations. This is because neither is the sole driver of the step in the model and so assuming their contamination to be zero is not valid. In fact, given that there are multiple drivers that are related but distinct from each other, there is no one-dimensional space in which the SN populations can be perfectly separated. It follows that for such a model, a single $\gamma_X$ for any tracer $X$ can never remove the full step.

\section{Discussion}
\label{sec:discussion}

Identifying the underlying driver of the SN Ia luminosity step is  important for precision cosmological measurements, as well as in the physical understanding of SN Ia explosions. This driver is unlikely to be a direct observable (e.g., a magnitude or colour) or a simple physical property that can be directly inferred from observations (e.g., global stellar mass), hence the search for the best indirect tracer has seen increasing focus in recent work. Here we discuss the implications of our simulations in two aspects: first, we discuss the contamination of tracers and their relation to physical parameters, and then we discuss our work in the context of cosmological distance measurements.

\subsection{What drives the step and what is the best tracer of it? }

\citetalias{Briday2021} find that spectroscopic local sSFR is the best environmental tracer of the SN Ia luminosity step: using it as a reference tracer minimises the dispersion around the best-fitting step versus contamination relationship. The local sSFR measurement itself is unlikely to relate directly to SN Ia luminosities. Instead, local sSFR presumably acts as a tracer of some fundamental underlying properties of the local stellar population or SN Ia progenitor. These might include their respective ages, which trace relationships between galaxy evolution and feedback, dust creation and destruction, metallicity, and white dwarf mass, which all have differing levels of importance. By testing purely SN Ia progenitor-age-driven models (Section \ref{subsec:res_age_step_only}), galaxy-age-driven models (Section \ref{subsubsec:res_age_rv_step_only}), and a combination of the two (Section \ref{subsubsec:res_ageplusstep}), we have been able to test both the underlying drivers and how well they are traced by measurable parameters. 

\subsubsection{Contamination}

We first consider tracer contamination. The contamination measurements presented in \citetalias{Briday2021} are derived directly from observations but are only relative and depend upon tracer choices, while those measured in our simulations are connected directly to step drivers but depend upon the relationships built into the galaxy evolution model of \citet{Wiseman2021}. By comparing both observed and simulated contamination for given observables against our synthetic reference tracers, no tracer particularly closely follows the SN progenitor age (see left-hand panels of Figs.~\ref{fig:SNage_step_only}--\ref{fig:Rv+SNage}). Galaxy age is 55 per cent contaminated with reference to SN progenitor age. This is because of the declining power-law nature of the DTD: roughly half of SNe Ia explode within 1\,Gyr of progenitor system formation and half beyond 1\,Gyr. This smooths out the relationship between the age of a stellar population and the SNe Ia that explode within it. The best tracer (of all observed and simulated) of SN Ia progenitor age is spectroscopic local sSFR, as identified in \citet{Rigault2018} and \citetalias{Briday2021}. Interestingly it is a better tracer of SN progenitor age than the overall galaxy age: the fact that the measurement is local carries more information about the progenitor than is lost by using a tracer (sSFR) instead of knowing the property directly. 

When choosing galaxy age as a reference instead of SN progenitor age, the measurable properties of spectroscopic local sSFR, (local) galaxy colours, and stellar mass are much less contaminated. This is because they are more closely related to the age of the stellar population than they are to an individual SN progenitor age. Spectroscopic local sSFR can still be used to broadly trace SN progenitor age -- spectroscopic local sSFR separates the SN Ia stretch population well \citep{Nicolas2020}, and SN stretch is much more likely to be linked to progenitor age than galaxy age. Nevertheless, our results show that neither spectroscopic local sSFR nor any other tracer should be expected to separate progenitor populations based on their age any better than with $\geq30$~per cent contamination. On the other hand, our results indicate that spectroscopic local sSFR can separate stellar population ages with $\simeq 10$~per cent contamination. At low redshift, where galaxies can be more easily resolved and this tracer is therefore measurable, we encourage observations to obtain spectroscopic local sSFR in SN host galaxies in order to trace the step driver accurately. 

At higher redshifts where it is not currently possible to obtain spectroscopy with 1\,kpc resolution, local photometric and global measurements are the only available tracers. In such cases, if it is possible to measure local galaxy colour then this property traces either SN progenitor age or galaxy age best. At high redshift (roughly $z \gtrsim 0.6$ in DES, although this limit will be higher with better seeing in the Legacy Survey of Space and Time; \citealt{Ivezic2019}), where the imaging cannot resolve local regions smaller than a few kpc, local measurements of colour and sSFR are not possible. In this case, the specific choice of global tracer makes little difference when tracing SN progenitor age. 

Photometric sSFR is marginally disfavoured when tracing galaxy age. This result itself is slightly counter-intuitive because sSFR is usually assumed to be a strong discriminator of galaxy SFHs and thus ages. We find the large contamination is due to the relatively broad range of stellar mass at which galaxy quenching can occur in our simulations, which was chosen to best reproduce observations of galaxy populations. This range translates to a range of ages at which quenching occurs, hence blurring out the correlation.

The lack of range between the contamination of the simulated tracers compared to the much larger respective range for the observed data is likely a combination of two factors. Firstly, the observations are local measurements, while the simulation does not carry spatial information so only global measurements are possible. Future work will improve the model to include spatial dimensions in the galaxy evolution model. A second reason for the lack of range is that the simulations suffer from oversimplifications that lead to missing or washed-out relationships, or a lack of diversity in the galaxy population. Again, these issues will be addressed in future versions of the simulations.

\subsubsection{The true driver of the step}

So far we have shown that measurable galaxy properties are better tracers of galaxy age than of SN progenitor age, but that doesn't necessarily mean that galaxy age is the true step driver. Instead, a model representing the true driver should result in a tight correlation between step magnitudes and contamination for each tracer. 
In Section \ref{subsec:res_age_rv_steps} we find a good match between the simulated and measured steps and tracer contamination assuming a model with a large $R_V$ difference and no intrinsic luminosity step, using galaxy age as the reference tracer. Such a result indicates that a galaxy-age varying $R_V$ is consistent with causing the full step independent of the choice of tracer and that there is no need for an additional difference in the luminosity of SNe as a function of their age. The model with a small $R_V$ difference plus an intrinsic luminosity step shows a worse agreement than the no-intrinsic step model, using either SN progenitor age or galaxy age as reference tracers. Of these, SN progenitor age provides a better match but with all other tracers displaying steps larger than predicted.

Taken at face value this would add further weight to the argument for a single galaxy-age $R_V$ step as the sole driver. The reality is less clear cut: in the model with both $R_V$ and intrinsic steps, we know that neither SN progenitor age nor galaxy age are the sole true driver since the luminosity varies as a combination of the two. Thus, neither has a true contamination of zero rendering the step--contamination diagram inaccurate. We thus do not rule out that model as being the true driver of the observed steps, but that a one-dimensional step (or two-dimensional contamination--step parameter space) is an inadequate discriminator for a model with multiple drivers of luminosity variations. It has been shown in previous work that a combination of environmental tracers can provide a better standardization than a single tracer \citep[e.g.,][]{Rose2021,Kelsey2022}. The development of a statistical discriminator between multi-driver models is left for future work. 

\subsection{Impact on cosmological measurements}
Since their discovery \citep{Kelly2010,Lampeitl2010,Sullivan2010}, host galaxy--SN luminosity correlations have usually been removed from SN Ia distance measurements via a single standardization parameter, $\gamma$ (Eq.~\ref{eq:tripp}). In parallel, the last decade has seen vast improvements in the methodology for correcting Malmquist-like biases, or selection biases, in SN Ia samples. From simple redshift-only corrections \citep[e.g.,][]{Betoule2014} the frameworks are now in place to perform bias corrections in many dimensions, including based upon parameters of the SN light curves and host galaxies \citep{Scolnic2016,Popovic2021a,Popovic2021}. At this point, the standardization parameters and bias correction routines become intricately connected. For example, correcting for a bias involving the light-curve stretch is also host galaxy dependent (because of stretch--host galaxy relationships) and thus affects the measured value of $\gamma$ \citep{Smith2020}. 

A recent bias correction model, \lq BBC4D\rq\ \citep{Popovic2021}, is based directly on the $R_V$ models used in this work. BBC4D corrects for dust-related luminosity bias by assuming different values of $R_V$ in hosts below and above a threshold in some tracer, nominally stellar mass. The results of this work demonstrate that if galaxy age is the true driver of the $R_V$ difference, populations split by stellar mass are nearly 40 per cent contaminated, meaning using stellar mass as the host parameter in BBC4D may not fully account for the bias/step. On the other hand, we show that other global measurements, particularly host colour, are no less contaminated. We thus predict, assuming galaxy age as the driver, that SN distances inferred using BBC4D may still display a residual luminosity step or may not have their intrinsic scatter reduced as much as would be achievable with a less contaminated measurement such as spectroscopic local sSFR. Similar, if not worse, biases are expected for an analysis not implementing $R_V$-based bias corrections but using simple $\gamma$ nuisance parameters. Future work will focus on estimating and reducing the bias on cosmological measurements introduced by tracer contamination.

\section{Conclusion}
\label{sec:conclusions}

In this work we have investigated the astrophysical driver of the SN Ia environmental luminosity step, by comparing the size of steps inferred by different environmental tracers in a simulated sample of SNe Ia, and how contaminated those tracers are with respect to the driver of the step. We first validated the method by inputting a simple SN-progenitor-age -- luminosity step and testing the resulting relationship between the measured step and contamination of each environmental tracer. We then tested a model where the entire step is caused by a large difference in dust extinction slope $R_V$ between young and old galaxies, and one where the $R_V$ difference is smaller and there is an additional, intrinsic step in SN luminosity as a function of the SN progenitor age. We find:
\begin{enumerate}
    \item in our simulation of a simple SN progenitor age step, we recover the expected linear relationship between step size and tracer contamination which matches the trend seen in observations.
    \item modelling SNe Ia with an $R_V$ that changes with host galaxy age as the sole driver of luminosity differences, and fixing galaxy age as the reference tracer, results in the best correlation between step size and contamination.
    \item local sSFR is less contaminated when assuming galaxy age as a reference tracer than when SN progenitor age is the reference.
    \item that the step versus contamination relationship is looser for a model with both a galaxy-age -- $R_V$ step and an intrinsic SN luminosity step, regardless of whether the reference is SN progenitor age or galaxy age.
    \item that the simple tracer contamination versus step parameter space is not adequate to constrain models with multiple parameters driving the step.
    \item that the implication of point (v) is that using a single $\gamma$ or a single tracer in a host-dependent bias correction will not remove the full effect of the step.

\end{enumerate}
The choice of tracer is deeply woven into the measurement of cosmological parameters. At face value, we confirm the result of \citetalias{Briday2021} that spectroscopic local sSFR is the closest representation of the driver and should be used to correct distances where available. At high redshift, where that measurement is not possible, both stellar mass and $U-R$ colour are the best, non-ideal, tracers. We caution against performing cosmological measurements using a simplistic $\gamma$ correction based on a single tracer and encourage the modelling of the step alongside selection effects.

\section*{Acknowledgements}

PW acknowledges support from the Science and Technology
Facilities Council (STFC) grant ST/R000506/1.

\section*{Data Availability}
 
This work is based on simulations from publicly available code (\url{https://github.com/wisemanp/des_sn_hosts}) and data from a published paper (\citetalias{Briday2021}).



\bibliographystyle{mnras}
\bibliography{PhilMendeley2} 

\begin{thebibliography}{}
\makeatletter
\relax
\def\mn@urlcharsother{\let\do\@makeother \do\$\do\&\do\#\do\^\do\_\do\%\do\~}
\def\mn@doi{\begingroup\mn@urlcharsother \@ifnextchar [ {\mn@doi@}
  {\mn@doi@[]}}
\def\mn@doi@[#1]#2{\def\@tempa{#1}\ifx\@tempa\@empty \href
  {http://dx.doi.org/#2} {doi:#2}\else \href {http://dx.doi.org/#2} {#1}\fi
  \endgroup}
\def\mn@eprint#1#2{\mn@eprint@#1:#2::\@nil}
\def\mn@eprint@arXiv#1{\href {http://arxiv.org/abs/#1} {{\tt arXiv:#1}}}
\def\mn@eprint@dblp#1{\href {http://dblp.uni-trier.de/rec/bibtex/#1.xml}
  {dblp:#1}}
\def\mn@eprint@#1:#2:#3:#4\@nil{\def\@tempa {#1}\def\@tempb {#2}\def\@tempc
  {#3}\ifx \@tempc \@empty \let \@tempc \@tempb \let \@tempb \@tempa \fi \ifx
  \@tempb \@empty \def\@tempb {arXiv}\fi \@ifundefined
  {mn@eprint@\@tempb}{\@tempb:\@tempc}{\expandafter \expandafter \csname
  mn@eprint@\@tempb\endcsname \expandafter{\@tempc}}}

\bibitem[\protect\citeauthoryear{Abbott et~al.,}{Abbott
  et~al.}{2019}]{Abbott2019}
Abbott T. M.~C.,  et~al., 2019, \mn@doi [ApJ] {10.3847/2041-8213/ab04fa}, 872,
  L30

\bibitem[\protect\citeauthoryear{Bernstein et~al.,}{Bernstein
  et~al.}{2012}]{Bernstein2012}
Bernstein J.~P.,  et~al., 2012, \mn@doi [ApJ] {10.1088/0004-637X/753/2/152},
  753, 152

\bibitem[\protect\citeauthoryear{Betoule et~al.,}{Betoule
  et~al.}{2014}]{Betoule2014}
Betoule M.,  et~al., 2014, \mn@doi [A\&A] {10.1051/0004-6361/201423413}, 568,
  A22

\bibitem[\protect\citeauthoryear{Briday et~al.,}{Briday
  et~al.}{2022}]{Briday2021}
Briday M.,  et~al., 2022, \mn@doi [A\&A] {10.1051/0004-6361/202141160}, 657,
  A22

\bibitem[\protect\citeauthoryear{Brout \& Scolnic}{Brout \&
  Scolnic}{2020}]{Brout2020}
Brout D.,  Scolnic D.,  2020, \mn@doi [ApJ] {10.3847/1538-4357/abd69b}, 909, 26

\bibitem[\protect\citeauthoryear{Brout et~al.,}{Brout et~al.}{2022}]{Brout2022}
Brout D.,  et~al., 2022, \mn@doi [ApJ] {10.3847/1538-4357/ac8e04}, 938, 110

\bibitem[\protect\citeauthoryear{Chen et~al.,}{Chen et~al.}{2022}]{Chen2022}
Chen R.,  et~al., 2022, \mn@doi [ApJ] {10.3847/1538-4357/ac8b82}, 938, 62

\bibitem[\protect\citeauthoryear{Childress et~al.,}{Childress
  et~al.}{2013}]{Childress2013a}
Childress M.,  et~al., 2013, \mn@doi [ApJ] {10.1088/0004-637X/770/2/108}, 770,
  108

\bibitem[\protect\citeauthoryear{Childress, Wolf  \& Zahid}{Childress
  et~al.}{2014}]{Childress2014}
Childress M.~J.,  Wolf C.,   Zahid H.~J.,  2014, \mn@doi [MNRAS]
  {10.1093/mnras/stu1892}, 445, 1898

\bibitem[\protect\citeauthoryear{Duarte et~al.,}{Duarte
  et~al.}{2022}]{Duarte2022}
Duarte J.,  et~al., 2022, eprint arXiv:2211.14291

\bibitem[\protect\citeauthoryear{Gallagher, Garnavich, Caldwell, Kirshner, Jha,
  Li, Ganeshalingam  \& Filippenko}{Gallagher et~al.}{2008}]{Gallagher2008}
Gallagher J.~S.,  Garnavich P.~M.,  Caldwell N.,  Kirshner R.~P.,  Jha S.~W.,
  Li W.,  Ganeshalingam M.,   Filippenko A.~V.,  2008, \mn@doi [ApJ]
  {10.1086/590659}, 685, 752

\bibitem[\protect\citeauthoryear{Gonz{\'{a}}lez-Gait{\'{a}}n, de Jaeger,
  Galbany, Mour{\~{a}}o, Paulino-Afonso  \&
  Filippenko}{Gonz{\'{a}}lez-Gait{\'{a}}n et~al.}{2021}]{Gonzalez-Gaitan2020}
Gonz{\'{a}}lez-Gait{\'{a}}n S.,  de Jaeger T.,  Galbany L.,  Mour{\~{a}}o A.,
  Paulino-Afonso A.,   Filippenko A.~V.,  2021, \mn@doi [MNRAS]
  {10.1093/mnras/stab2802}, 508, 4656

\bibitem[\protect\citeauthoryear{Guy et~al.,}{Guy et~al.}{2007}]{Guy2007}
Guy J.,  et~al., 2007, \mn@doi [A\&A] {10.1051/0004-6361:20066930}, 466, 11

\bibitem[\protect\citeauthoryear{Guzm{\'{a}}n, Gallego, Koo, Phillips,
  Lowenthal, Faber, Illingworth  \& Vogt}{Guzm{\'{a}}n
  et~al.}{1997}]{Guzman1997}
Guzm{\'{a}}n R.,  Gallego J.,  Koo D.~C.,  Phillips A.~C.,  Lowenthal J.~D.,
  Faber S.~M.,  Illingworth G.~D.,   Vogt N.~P.,  1997, \mn@doi [ApJ]
  {10.1086/304797}, 489, 559

\bibitem[\protect\citeauthoryear{Hakobyan, Barkhudaryan, Karapetyan, Gevorgyan,
  Mamon, Kunth, Adibekyan  \& Turatto}{Hakobyan et~al.}{2020}]{Hakobyan2020}
Hakobyan A.~A.,  Barkhudaryan L.~V.,  Karapetyan A.~G.,  Gevorgyan M.~H.,
  Mamon G.~A.,  Kunth D.,  Adibekyan V.,   Turatto M.,  2020, \mn@doi [MNRAS]
  {10.1093/mnras/staa2940}, 499, 1424

\bibitem[\protect\citeauthoryear{Hamuy, Phillips, Maza, Suntzeff, Schommer  \&
  Aviles}{Hamuy et~al.}{1995}]{Hamuy1995}
Hamuy M.,  Phillips M.~M.,  Maza J.,  Suntzeff N.~B.,  Schommer R.~A.,   Aviles
  R.,  1995, \mn@doi [Astron. J.] {10.1086/117251}, 109, 1

\bibitem[\protect\citeauthoryear{Ivezi{\'{c}} et~al.,}{Ivezi{\'{c}}
  et~al.}{2019}]{Ivezic2019}
Ivezi{\'{c}} {\v{Z}}.,  et~al., 2019, \mn@doi [ApJ] {10.3847/1538-4357/ab042c},
  873, 111

\bibitem[\protect\citeauthoryear{Johansson et~al.,}{Johansson
  et~al.}{2021}]{Johansson2021}
Johansson J.,  et~al., 2021, \mn@doi [ApJ] {10.3847/1538-4357/ac2f9e}, 923, 237

\bibitem[\protect\citeauthoryear{Jones et~al.,}{Jones et~al.}{2018}]{Jones2018}
Jones D.~O.,  et~al., 2018, \mn@doi [ApJ] {10.3847/1538-4357/aae2b9}, 867, 108

\bibitem[\protect\citeauthoryear{Jones et~al.,}{Jones et~al.}{2022}]{Jones2022}
Jones D.~O.,  et~al., 2022, \mn@doi [ApJ] {10.3847/1538-4357/ac755b}, 933, 172

\bibitem[\protect\citeauthoryear{Kelly, Hicken, Burke, Mandel  \&
  Kirshner}{Kelly et~al.}{2010}]{Kelly2010}
Kelly P.~L.,  Hicken M.,  Burke D.~L.,  Mandel K.~S.,   Kirshner R.~P.,  2010,
  \mn@doi [ApJ] {10.1088/0004-637X/715/2/743}, 715, 743

\bibitem[\protect\citeauthoryear{Kelsey et~al.,}{Kelsey
  et~al.}{2021}]{Kelsey2021}
Kelsey L.,  et~al., 2021, \mn@doi [MNRAS] {10.1093/mnras/staa3924}, 501, 4861

\bibitem[\protect\citeauthoryear{Kelsey et~al.,}{Kelsey
  et~al.}{2022}]{Kelsey2022}
Kelsey L.,  et~al., 2022, \mn@doi [MNRAS] {10.1093/mnras/stac3711}, 519, 3046

\bibitem[\protect\citeauthoryear{Kessler et~al.,}{Kessler
  et~al.}{2009}]{Kessler2009}
Kessler R.,  et~al., 2009, \mn@doi [ApJS] {10.1088/0067-0049/185/1/32}, 185, 32

\bibitem[\protect\citeauthoryear{Kessler et~al.,}{Kessler
  et~al.}{2019}]{Kessler2019}
Kessler R.,  et~al., 2019, \mn@doi [MNRAS] {10.1093/mnras/stz463}, 485, 1171

\bibitem[\protect\citeauthoryear{Lampeitl et~al.,}{Lampeitl
  et~al.}{2010}]{Lampeitl2010}
Lampeitl H.,  et~al., 2010, \mn@doi [MNRAS] {10.1111/j.1365-2966.2009.15851.x},
  401, 2331

\bibitem[\protect\citeauthoryear{Maoz, Mannucci  \& Nelemans}{Maoz
  et~al.}{2014}]{Maoz2014}
Maoz D.,  Mannucci F.,   Nelemans G.,  2014, \mn@doi [Annu. Rev. Astron.
  Astrophys.] {10.1146/annurev-astro-082812-141031}, 52, 107

\bibitem[\protect\citeauthoryear{Meldorf et~al.,}{Meldorf
  et~al.}{2022}]{Meldorf2022}
Meldorf C.,  et~al., 2022, \mn@doi [MNRAS] {10.1093/mnras/stac3056}, 518, 1985

\bibitem[\protect\citeauthoryear{Mill{\'{a}}n-Irigoyen, del Valle-Espinosa,
  Fern{\'{a}}ndez-Aranda, Galbany, Gomes, Moreno-Raya,
  L{\'{o}}pez-S{\'{a}}nchez  \& Moll{\'{a}}}{Mill{\'{a}}n-Irigoyen
  et~al.}{2022}]{Millan-Irigoyen2022}
Mill{\'{a}}n-Irigoyen I.,  del Valle-Espinosa M.~G.,  Fern{\'{a}}ndez-Aranda
  R.,  Galbany L.,  Gomes J.~M.,  Moreno-Raya M.,  L{\'{o}}pez-S{\'{a}}nchez
  {\'{A}}.~R.,   Moll{\'{a}} M.,  2022, \mn@doi [MNRAS]
  {10.1093/mnras/stac2696}, 517, 3312

\bibitem[\protect\citeauthoryear{Moreno-Raya, Moll{\'{a}},
  L{\'{o}}pez-S{\'{a}}nchez, Galbany, V{\'{i}}lchez, Rosell  \&
  Dom{\'{i}}nguez}{Moreno-Raya et~al.}{2016}]{Moreno-Raya2016}
Moreno-Raya M.~E.,  Moll{\'{a}} M.,  L{\'{o}}pez-S{\'{a}}nchez {\'{A}}.~R.,
  Galbany L.,  V{\'{i}}lchez J.~M.,  Rosell A.~C.,   Dom{\'{i}}nguez I.,  2016,
  \mn@doi [ApJ] {10.3847/2041-8205/818/1/l19}, 818, L19

\bibitem[\protect\citeauthoryear{Nicolas et~al.,}{Nicolas
  et~al.}{2021}]{Nicolas2020}
Nicolas N.,  et~al., 2021, \mn@doi [A\&A] {10.1051/0004-6361/202038447}, 649,
  A74

\bibitem[\protect\citeauthoryear{Perlmutter et~al.,}{Perlmutter
  et~al.}{1999}]{Perlmutter1999}
Perlmutter S.,  et~al., 1999, \mn@doi [ApJ] {10.1086/307221}, 517, 565

\bibitem[\protect\citeauthoryear{Perrett et~al.,}{Perrett
  et~al.}{2010}]{Perrett2010}
Perrett K.,  et~al., 2010, \mn@doi [Astron. J.] {10.1088/0004-6256/140/2/518},
  140, 518

\bibitem[\protect\citeauthoryear{Phillips}{Phillips}{1993}]{Phillips1993}
Phillips M.~M.,  1993, \mn@doi [ApJ] {10.1086/186970}, 413, L105

\bibitem[\protect\citeauthoryear{Ponder, Wood-Vasey, Weyant, Barton, Galbany,
  Liu, Garnavich  \& Matheson}{Ponder et~al.}{2021}]{Ponder2020}
Ponder K.~A.,  Wood-Vasey W.~M.,  Weyant A.,  Barton N.~T.,  Galbany L.,  Liu
  S.,  Garnavich P.,   Matheson T.,  2021, \mn@doi [ApJ]
  {10.3847/1538-4357/ac2d99}, 923, 197

\bibitem[\protect\citeauthoryear{Popovic, Brout, Kessler  \& Scolnic}{Popovic
  et~al.}{2021a}]{Popovic2021a}
Popovic B.,  Brout D.,  Kessler R.,   Scolnic D.,  2021a, eprint
  arxiv:2112.04456

\bibitem[\protect\citeauthoryear{Popovic, Brout, Kessler, Scolnic  \&
  Lu}{Popovic et~al.}{2021b}]{Popovic2021}
Popovic B.,  Brout D.,  Kessler R.,  Scolnic D.,   Lu L.,  2021b, \mn@doi [ApJ]
  {10.3847/1538-4357/abf14f}, 913, 49

\bibitem[\protect\citeauthoryear{Riess, Press  \& Kirshner}{Riess
  et~al.}{1996}]{Riess1996}
Riess A.~G.,  Press W.~H.,   Kirshner R.~P.,  1996, \mn@doi [ApJ]
  {10.1086/178129}, 473, 88

\bibitem[\protect\citeauthoryear{Riess et~al.,}{Riess et~al.}{1998}]{Riess1998}
Riess A.~G.,  et~al., 1998, \mn@doi [Astron. J.] {10.1086/300499}, 116, 1009

\bibitem[\protect\citeauthoryear{Rigault et~al.,}{Rigault
  et~al.}{2013}]{Rigault2013}
Rigault M.,  et~al., 2013, \mn@doi [A\&A] {10.1051/0004-6361/201322104}, 560,
  A66

\bibitem[\protect\citeauthoryear{Rigault et~al.,}{Rigault
  et~al.}{2020}]{Rigault2018}
Rigault M.,  et~al., 2020, \mn@doi [A\&A] {10.1051/0004-6361/201730404}, 644,
  A176

\bibitem[\protect\citeauthoryear{Roman et~al.,}{Roman et~al.}{2018}]{Roman2018}
Roman M.,  et~al., 2018, \mn@doi [A\&A] {10.1051/0004-6361/201731425}, 615, A68

\bibitem[\protect\citeauthoryear{Rose, Garnavich  \& Berg}{Rose
  et~al.}{2019}]{Rose2019}
Rose B.~M.,  Garnavich P.~M.,   Berg M.~A.,  2019, \mn@doi [ApJ]
  {10.3847/1538-4357/ab0704}, 874, 32

\bibitem[\protect\citeauthoryear{Rose, Rubin, Strolger  \& Garnavich}{Rose
  et~al.}{2021}]{Rose2021}
Rose B.~M.,  Rubin D.,  Strolger L.,   Garnavich P.~M.,  2021, \mn@doi [ApJ]
  {10.3847/1538-4357/abd550}, 909, 28

\bibitem[\protect\citeauthoryear{Salim, Boquien  \& Lee}{Salim
  et~al.}{2018}]{Salim2018}
Salim S.,  Boquien M.,   Lee J.~C.,  2018, \mn@doi [ApJ]
  {10.3847/1538-4357/aabf3c}, 859, 11

\bibitem[\protect\citeauthoryear{Scolnic \& Kessler}{Scolnic \&
  Kessler}{2016}]{Scolnic2016}
Scolnic D.,  Kessler R.,  2016, \mn@doi [ApJ] {10.3847/2041-8205/822/2/L35},
  822, L35

\bibitem[\protect\citeauthoryear{Scolnic et~al.,}{Scolnic
  et~al.}{2018}]{Scolnic2018}
Scolnic D.~M.,  et~al., 2018, \mn@doi [ApJ] {10.3847/1538-4357/aab9bb}, 859,
  101

\bibitem[\protect\citeauthoryear{Smith et~al.,}{Smith et~al.}{2020}]{Smith2020}
Smith M.,  et~al., 2020, \mn@doi [MNRAS] {10.1093/mnras/staa946}, 494, 4426

\bibitem[\protect\citeauthoryear{Sullivan et~al.,}{Sullivan
  et~al.}{2003}]{Sullivan2003}
Sullivan M.,  et~al., 2003, \mn@doi [MNRAS] {10.1046/j.1365-8711.2003.06312.x},
  340, 1057

\bibitem[\protect\citeauthoryear{Sullivan et~al.,}{Sullivan
  et~al.}{2006}]{Sullivan2006}
Sullivan M.,  et~al., 2006, \mn@doi [ApJ] {10.1086/506137}, 648, 868

\bibitem[\protect\citeauthoryear{Sullivan et~al.,}{Sullivan
  et~al.}{2010}]{Sullivan2010}
Sullivan M.,  et~al., 2010, \mn@doi [MNRAS] {10.1111/j.1365-2966.2010.16731.x},
  406, 782

\bibitem[\protect\citeauthoryear{Sullivan et~al.,}{Sullivan
  et~al.}{2011}]{Sullivan2011}
Sullivan M.,  et~al., 2011, \mn@doi [ApJ] {10.1088/0004-637X/737/2/102}, 737,
  102

\bibitem[\protect\citeauthoryear{Thorp \& Mandel}{Thorp \&
  Mandel}{2022}]{Thorp2022}
Thorp S.,  Mandel K.~S.,  2022, \mn@doi [MNRAS] {10.1093/mnras/stac2714}, 517,
  2360

\bibitem[\protect\citeauthoryear{Thorp, Mandel, Jones, Ward  \& Narayan}{Thorp
  et~al.}{2021}]{Thorp2021}
Thorp S.,  Mandel K.~S.,  Jones D.~O.,  Ward S.~M.,   Narayan G.,  2021,
  \mn@doi [MNRAS] {10.1093/mnras/stab2849}, 508, 4310

\bibitem[\protect\citeauthoryear{Tripp}{Tripp}{1998}]{Tripp1998}
Tripp R.,  1998, A\&A, 331, 815

\bibitem[\protect\citeauthoryear{Uddin et~al.,}{Uddin et~al.}{2020}]{Uddin2020}
Uddin S.~A.,  et~al., 2020, \mn@doi [ApJ] {10.3847/1538-4357/ABAFB7}, 901, 143

\bibitem[\protect\citeauthoryear{Vincenzi et~al.,}{Vincenzi
  et~al.}{2021}]{Vincenzi2020}
Vincenzi M.,  et~al., 2021, \mn@doi [MNRAS] {10.1093/mnras/stab1353}, 505, 2819

\bibitem[\protect\citeauthoryear{Wiseman et~al.,}{Wiseman
  et~al.}{2021}]{Wiseman2021}
Wiseman P.,  et~al., 2021, \mn@doi [MNRAS] {10.1093/mnras/stab1943}, 506, 3330

\bibitem[\protect\citeauthoryear{Wiseman et~al.,}{Wiseman
  et~al.}{2022}]{Wiseman2022}
Wiseman P.,  et~al., 2022, \mn@doi [MNRAS] {10.1093/mnras/stac1984}, 515, 4587

\makeatother
\end{thebibliography}





\bsp	
\label{lastpage}
\end{document}